\documentclass[prd,twocolumn,nofootinbib,showpacs,preprintnumbers,floatfix]{revtex4}
\usepackage[plainpages=false, colorlinks=true, anchorcolor=blue, linkcolor=blue, citecolor=blue, bookmarks=false]{hyperref}
\usepackage[utf8]{inputenc}
\usepackage{amsfonts,amsmath,amssymb}
\usepackage{graphicx}
\usepackage{color}
\usepackage{natbib}
\usepackage{enumitem}
\usepackage{subcaption}
\usepackage{comment}
\newcommand{\rthis}[1]{\textcolor{black}{#1}}
\begin{document}
\newcommand{\apjl}{Astrophys. J. Lett.}
\newcommand{\apjs}{Astrophys. J. Suppl. Ser.}
\newcommand{\aap}{Astron. \& Astrophys.}
\newcommand{\aj}{Astron. J.}
\newcommand{\araa}{Ann. Rev. Astron. Astrophys. } 
\newcommand{\mnras}{Mon. Not. R. Astron. Soc.}
\newcommand{\apss} {Astrophys. and Space Science}
\newcommand{\jcap}{JCAP}
\newcommand{\pasj}{PASJ}
\newcommand{\pasa}{Pub. Astro. Soc. Aust.}
\newcommand{\physrep}{Physics Reports}
\title{Model Comparison of $\Lambda$CDM vs $R_h=ct$ using Cosmic Chronometers}
\author{Haveesh \surname{Singirikonda}}\altaffiliation{E-mail:ep17btech11010@iith.ac.in}

\author{Shantanu  \surname{Desai}}  
\altaffiliation{E-mail: shntn05@gmail.com}
\begin{abstract}
In 2012, Bilicki and Seikel~\cite{Seikel_2012_rhct} showed that  $H(z)$ data reconstructed using Gaussian Process Regression from cosmic chronometers and baryon acoustic oscillations,  conclusively rules out the $R_h=ct$ model. These results were disputed by Melia and collaborators in two different works~\cite{Maier,Meliachrono}, who showed using both an  unbinned analysis and Gaussian Process reconstructed  $H(z)$ data from chronometers, that  $R_h=ct$
is favored over $\Lambda$CDM model. To resolve this imbroglio, we carry out model comparison of $\Lambda$CDM versus $R_h=ct$ by independently reproducing the above claims using the latest chronometer data. We perform model selection between
these two models using  Bayesian model comparison. We find that  no one model between $\Lambda$CDM and  $R_h=ct$ is decisively favored \rthis{when uniform priors on $\Lambda$CDM parameters are used. However, if we use priors centered around the Planck best-fit values, then $\Lambda$CDM  is very strongly preferred over $R_h=ct$.}
\end{abstract}

\affiliation{Department of Physics, Indian Institute of Technology, Hyderabad, Telangana-502285, India}

\maketitle

\section{Introduction}
The standard hot Big-Bang model of cosmology is described by  a flat $\Lambda$CDM universe, with 70\% of the energy density comprising of the cosmological constant (or any dark energy fluid with equation of state $w \equiv P / \rho$ \rthis{close to -1)} and 25\% cold (non-baryonic) dark matter and 5\% baryons~\cite{Ratra03}. This model has two episodes of acceleration (one in the early universe caused by inflation~\cite{inflation}, posited to solve the horizon and flatness problems in the standard hot Big-Bang model~\cite{Dicke79}), and another in the late universe, caused by dark energy~\cite{Huterer}. This model has been spectacularly confirmed by Planck 2018 CMB observations~\cite{Planck2018} along with other large-scale structure probes. There are however a few data-driven lingering problems with the standard $\Lambda$CDM paradigm, such as the Hubble constant tension between local and high redshift measurements~\cite{Verde,Bethapudi}, $\sigma_8$ tension between CMB and galaxy clusters~\cite{Planck2016clusters,Bocquet}, Lithium-7 problem in Big-Bang nucleosynthesis~\cite{Fields}, anomalies in CMB at low $l$~\cite{Copi}, etc. A few works have also  challenged some of the most well-established tenets of the standard cosmological model,  viz. cosmic acceleration~\cite{Sarkar} and even cosmic expansion~\cite{expand}.

Independent of  the above data driven problems, there are also conceptual problems with the standard model. The best-fit model of scalar-field driven inflation (an essential pillar of standard hot Big-Bang model) with flat potentials also causes lots of fine-tuning issues~\cite{Loeb}. Furthermore,  we don't yet  have  laboratory evidence for any cold dark matter candidate, despite searching for over three decades~\cite{Merritt}. If the dark energy turns out to be a cosmological constant, a non-zero value  would be  very problematic from the point of view of quantum field theory~\cite{Weinberg,Martin}.

Therefore, because of some of the above problems, many alternatives to the standard model  have been constructed. One such model is the $R_h=ct$ universe model, proposed by Fulvio Melia~\cite{Melia07,Shevchuk,Melia2012}. In this model, the size of the Hubble sphere given by $R_h(t)=ct$ is upheld for all times in contrast to the case of the $\Lambda$CDM model, where this coincidence is true only at the current epoch, i.e.  $R_h(t_0)=ct_0$. This model has $a(t) \propto t$ and $H(z)=H_0(1+z)$.
One direct result of this is that the rate of expansion $\dot{a}$ is constant; and pressure and energy density satisfy an equation of state given by $p=-\frac{\rho}{3}$. \rthis{This is known as  the zero active mass condition, and has been argued by Melia to be a necessary requirement due to the symmetries of FRW universes~\cite{Melia2016}. (See however Ref.~\cite{Lasenby} for objections to this argument of zero active mass condition.)} Melia has also argued that this model provides a cosmological basis for the  origin of the rest mass energy relation, i.e. $E=mc^2$~\cite{Melia19}, although this has been disputed~\cite{Lewis19}.
The $R_h=ct$ model also has several antecedents and generalizations, discussed in Refs.~\cite{Moncy,Jain}, and an up-to-date  review of all such models can be found in Ref.~\cite{Casado}. This model has been tested with a whole slew of cosmological observations by Melia and collaborators; such as cosmic chronometers~\cite{Meliachrono}, quasar core angular size measurements~\cite{Meliaquasar}, quasar X-ray and UV fluxes~\cite{MeliaUV}, Type 1a SN~\cite{MeliaSN}, strong lensing~\cite{MeliaSL}, cluster gas mass fraction~\cite{Meliafgas}, etc  and found to be in better agreement compared to $\Lambda$CDM model. However, other researchers have reached opposite conclusions and have argued that this model is inconsistent  with observations~\cite{Shafer,Seikel_2012_rhct,LewisBBN,Haridasu1,Lin,Hu2018,Tu2019,Fuji}. \rthis{Even before this model was introduced, there were severe observational constraints on power-law cosmologies, within  which this model can be subsumed~\cite{Kaplinghat1,Kaplinghat2}.}
These results  in turn have also been contested by Melia and collaborators~\cite{Mcclintock}.
Conceptual problems have also been raised against this model~\cite{Kwan,Lewis12,Mitra,Lewishorizon,Lewisunphysical,Lasenby,Bengochea}, although some  have  been countered~\cite{Meliaamj}. We note however so far this model is yet to reproduce the Cosmic Microwave Background \rthis{temperature and polarization anisotropy measurements}.

In this work, we try to adjudicate between one such conflicting claim between two of the above works: Ref.~\cite{Seikel_2012_rhct} (BS12, hereafter) and Refs.~\cite{Maier,Meliachrono} (MM13 and MY18, hereafter),   which have reached diametrically opposite conclusions, when analyzing  Hubble parameter ($H(z)$) measurements. BS12 reconstructed a non-parametric fit for $H(z)$ using Gaussian Process Regression (GPR hereafter) from 18 cosmic chronometer measurements and 8 BAO measurements spanning the redshift range $0.09 \leq z \leq 0.73$. They argued based on a visual inspection of the reconstructed $H(z)$ and its derivatives, that the  $\Lambda$CDM model is a much better fit than the $R_h=ct$ model. Soon thereafter, MM13 however pointed out that  19 unbinned $H(z)$ measurements obtained from chronometers, support  $R_h=ct$ over the  $\Lambda$CDM. This assertion was based on AIC, BIC, and KIC based tests from information theory and $\chi^2$/dof. Most recently,  MY18 used 30 $H(z)$ measurements using cosmic chronometers, and similar to BS12, used GPR to reconstruct a non-parametric $H(z)$. Model comparison of $\Lambda$CDM vs $R_h=ct$ was done by calculating the normalized area difference between the model and the reconstructed $H(z)$. They argued that with this procedure, $R_h=ct$ model is a better fit than $\Lambda$CDM. Here, we  do an independent analysis of $H(z)$ data, using the latest  measurements from chronometers.

The outline of this paper is as follows. We discuss the GPR technique  and Bayesian model comparison technique  in Sect.~\ref{sec:GPR} and Sect.~\ref{sec:modelcomp} respectively. The key points made in the two conflicting sets of papers BS12 versus  MM13, MY18 are discussed in Sect.~\ref{sec:prevpapers}.  The description of our datasets and analysis can be found  in Sect.~\ref{sec:datasets}. Our results using $H(z)$ measurements can be found in Sect.~\ref{sec:results}. A comparison of the two models using the $Om(z_1,z_2)$ statistic can be found in Sect.~\ref{sec:om}. We  conclude in Sect.~\ref{sec:conclusions}.

\section{Gaussian Process Regression}
\label{sec:GPR}
Both the groups (BS12 and MY18) have used GPR for their analysis. Therefore, we provide an abridged introduction to GPR, before discussing the results of their analysis. A more detailed explanation can be found in Section 2 of Ref.~\cite{Seikel_2012}.
GPR is a widely used technique in astronomy as it allows us to smoothly interpolate in a non-parametric fashion between different datapoints, thereby allowing us to increase the number of degrees of freedom. However, they do not provide more information than the underlying data. Gaussian process is similar to a Gaussian distribution but it describes the distribution of functions instead of random variables. 
To describe the distribution of these functions, we need the mean function $\mu(x)$ and a covariance function $ cov(f(x),f(\Tilde{x})) = k(x,\Tilde{x}) $ connecting the values of $f$ evaluated at $x$ and $\Tilde{x}$. There are many choices for the covariance function. Both the papers have used a squared exponential/Gaussian covariance function, so even in this paper we use a Gaussian kernel for GPR. For a Gaussian kernel $k(x,\Tilde{x})$ is: $$ k(x,\Tilde{x}) = \sigma_f^2 \exp \left( -\frac{(x-\Tilde{x})^2}{2l^2} \right) $$ Here, $\sigma_f$ and $l$ are hyper-parameters which describe the `bumpiness' of the function. 

Even a random function $f(x)$ can be generated using the covariance matrix. Let $\mathbf{X}$ be the set of points ${x_i}$ and one can generate a vector $\mathbf{f^*}$ of function values at $\mathbf{X^*}$ with $f^*_i = f(x_i^*)$ as $$ \mathbf{f^*} = \mathcal{N}(\mathbf{\mu^*},K(\mathbf{X^*,X^*})) $$ The notation $\mathcal{N}$ means that  the Gaussian process is evaluated at $x^*$, where $f(x^*)$ is a random value drawn from a normal distribution. Similarly, observational data can be written in the same way as $$ \mathbf{y} = \mathcal{N}(\mathbf{\mu},K(\mathbf{X,X})+C) $$ where $C$ is the covariance matrix of the data. If data is uncorrelated the covariance matrix is simply $diag(\sigma_i^2)$. Using the values of $y$ at $\mathbf{X}$ we can reconstruct $\mathbf{f^*}$ using $$ \overline{\mathbf{f^*}} = \mathbf{\mu^*} + K(\mathbf{X^*,X})[K(\mathbf{X,X})+C]^{-1}(\mathbf{y-\mu}) $$ and $$ cov(\mathbf{f^*}) = K(\mathbf{X^*,X^*}) - K(\mathbf{X^*,X})[K(\mathbf{X,X})+C]^{-1}K(\mathbf{X,X^*}) $$ where $\overline{\mathbf{f^*}}$ and $cov(\mathbf{f^*})$ are mean and covariance of $\mathbf{f^*}$ respectively. The diagonal elements of $cov(\mathbf{f^*})$ provide us the variance of $\mathbf{f^*}$. More details on this can found in Ref.~\cite{Seikel_2012}. Both BS12 and MY18 implement GPR in {\tt Python} using the package {\tt GaPP}, which was developed by Seikel and collaborators~\cite{Seikel_2012}.

\section{Model Comparison summary}
\label{sec:modelcomp}

Model comparison between two models can  be broadly classified into three distinct categories: frequentist, information-theory, and Bayesian techniques~\cite{Liddle,Liddle_2007,Trotta,Shi,Weller}. 
In this work we shall only apply Bayesian model comparison, since this is argued to be the most robust among the different model comparison  techniques~\cite{Trotta,Sharma}. We briefly summarize this technique and more details can be found in Refs.~\cite{Trotta, Weller, Sharma} or some of our previous works~\cite{Krishak1,Krishak2}.

Using Bayesian statistics, we compute the probability that the data was generated by each model, also called the Bayesian evidence ($Z$)~\cite{Trotta}:
\begin{equation}
    P(\Theta|D,M) = \frac{P(D|\Theta,M) P(\Theta,M)}{P(D|M)}
    \label{eq:evidence}
\end{equation}
where $P(\Theta|D,M)$ is the posterior, $P(D|\Theta,M)$ is the likelihood, $P(\Theta,M)$ is the prior, and $P(D|M)$ is the evidence, also sometimes referred to as marginal likelihood. Note that unlike the other model comparison test, the Bayesian evidence does not use the best-fit value of a given model. It considers the entire range. Again, the model with a higher evidence, i.e, higher probability that the data was generated from that model, will be the better model to describe the data. From the   Bayesian evidence of the two models, we can calculate the value of the Bayes factor, which is simply the ratio of the evidence for the two models and given by:
\begin{equation}
    B = \frac{Z_1}{Z_2}.
\end{equation}

For the Bayes factor, we evaluate the ratio of the evidence of the  $\Lambda$CDM to the evidence for the $R_h=ct$ model. The significance can be evaluated using the Jeffreys scale~\cite{Trotta}.

\section{Summary of BS12, MM13, and  MY18} 
\label{sec:prevpapers}

As mentioned in the introduction, there is a large amount of literature  comparing the $R_h = ct$ model with the $\Lambda$CDM model. We focus on the particular case of these two sets of papers (BS12 versus MM13/MY18) and a few others which only use $H(z)$ measurements, where they have arrived at conflicting results despite  similar analysis.  We then briefly mention some other works which compared the two models using only expansion history.

BS12 reconstructed the value of  the deceleration parameter $q(z)$ from Union2.1 Type 1a Supernova dataset with GPR,
and showed from a visual inspection that the reconstructed $q(z)$ better fits the $\Lambda$CDM model.
They also used Hubble rate data from 18 cosmic chronometer  and 8 BAO measurements, and reconstructed $H(z)$ with GPR, and plotted it against the predicted values of $H(z)$ from the $\Lambda$CDM model and the $R_h = ct$ model. They compared  the reconstructed $H(z)$, its first and second derivative, as well as the $Om(z)$ diagnostic~\cite{Sahni:2008xx} against the theoretical predictions of the two models. They again used visual inspection from these plots to conclude  that the $\Lambda$CDM model is a better fit to the data compared to $R_h = ct$.
Very soon after BS12, MM13 considered 19 unbinned  $H(z)$ measurements from cosmic chronometers and fit this data to both the models. They found that the $\chi^2$/DOF (or reduced $\chi^2$) is equal to 0.745 and 0.777 for $R_h=ct$ and $\Lambda$CDM (with parameters given by: $\Omega_M=0.32$, $H_0= 68.9 \pm 2.4$ km/sec/Mpc) respectively. Therefore, the  reduced $\chi^2$ was smaller for $\Lambda$CDM.  However, when $\Lambda$CDM  model is fit to the cosmic chronometer data, the estimated values of $\Omega_M$ and $H_0$  (0.27 and $73.8 \pm 2.4$ km/sec/Mpc  respectively) yield a $\chi^2$/DOF of 0.9567, which is greater than that for $R_h=ct$
universe. However, no comparison of the goodness of fit based on $\chi^2$ p.d.f. was made. They also found smaller values of AIC, BIC, and KIC for $R_h=ct$ universe compared to $\Lambda$CDM.
However, we note that  the difference in information criterion between the two models did not cross the threshold of 10, needed for any one  model to be decisively favored over the other. They further criticized the SN data analysis in BS12, arguing that the data used was optimized for $\Lambda$CDM cosmology. They also argued  that the BAO data analyzed in BS12 includes non-linear evolution of the matter density and velocity fields, and hence is not model-independent. Therefore, their analysis was done using only chronometers.

A similar analysis using the latest cosmic chronometer data (consisting of 30 measurements) and GPR was carried out in MY18. Here, they  used an analytical approach to compare the two models after reconstructing the values of $H(z)$ using GPR. They argued  that the  $R_h=ct$ performs better than the $\Lambda$CDM model, contradicting the conclusion of BS12. To quantify this, they constructed  a mock data set using Gaussian random variables, and then computed the normalized absolute area difference between this and the real function. For each model they calculated the differential areas by replacing the mock data set with the predictions from the models, and then estimated the probability of the model ($p$-value). From this analysis they came to the conclusion that the $R_h=ct$ model is the better model  among the two for the chronometer data.  

Besides the above two sets of papers, Ref.~\cite{Lin} showed using AIC and BIC that a combination of JLA type 1a SN sample and 30 $H(z)$ measurements from chronometers and BAO strongly support the $\Lambda$CDM model over $R_h=ct$ universe. They also found using AIC and BIC that the chronometer only measurements by themselves do not  decisively favor any one model. 
Haridasu et al~\cite{Haridasu1}  did a joint analysis of Type 1a SN, BAO, GRB and chronometer $H(z)$ data and compared the likelihood of $\Lambda$CDM model with $R_h=ct$ using AIC and BIC. They found that both  $\Delta$AIC and $\Delta$BIC between the two models is greater than 20, thereby decisively ruling out  $R_h=ct$ model. \rthis{Hu and Wang showed from a test of the cosmic distance duality relation using a sample of galaxy clusters and Type 1a SN, that the $\Lambda$CDM model is strongly favored over $R_h=ct$ with both $\Delta$AIC and $\Delta$BIC greater than 10~\cite{Hu2018}. Tu et al used  a combination of strong lensing, Type 1a supernovae, BAO and  cosmic chronometers to argue that $\Lambda$CDM is moderately  favored over $R_h=ct$ model with the natural logarithm of the Bayes factor greater than five~\cite{Tu2019}.} 


\begin{figure}
\centering
\includegraphics[width=0.5\textwidth]{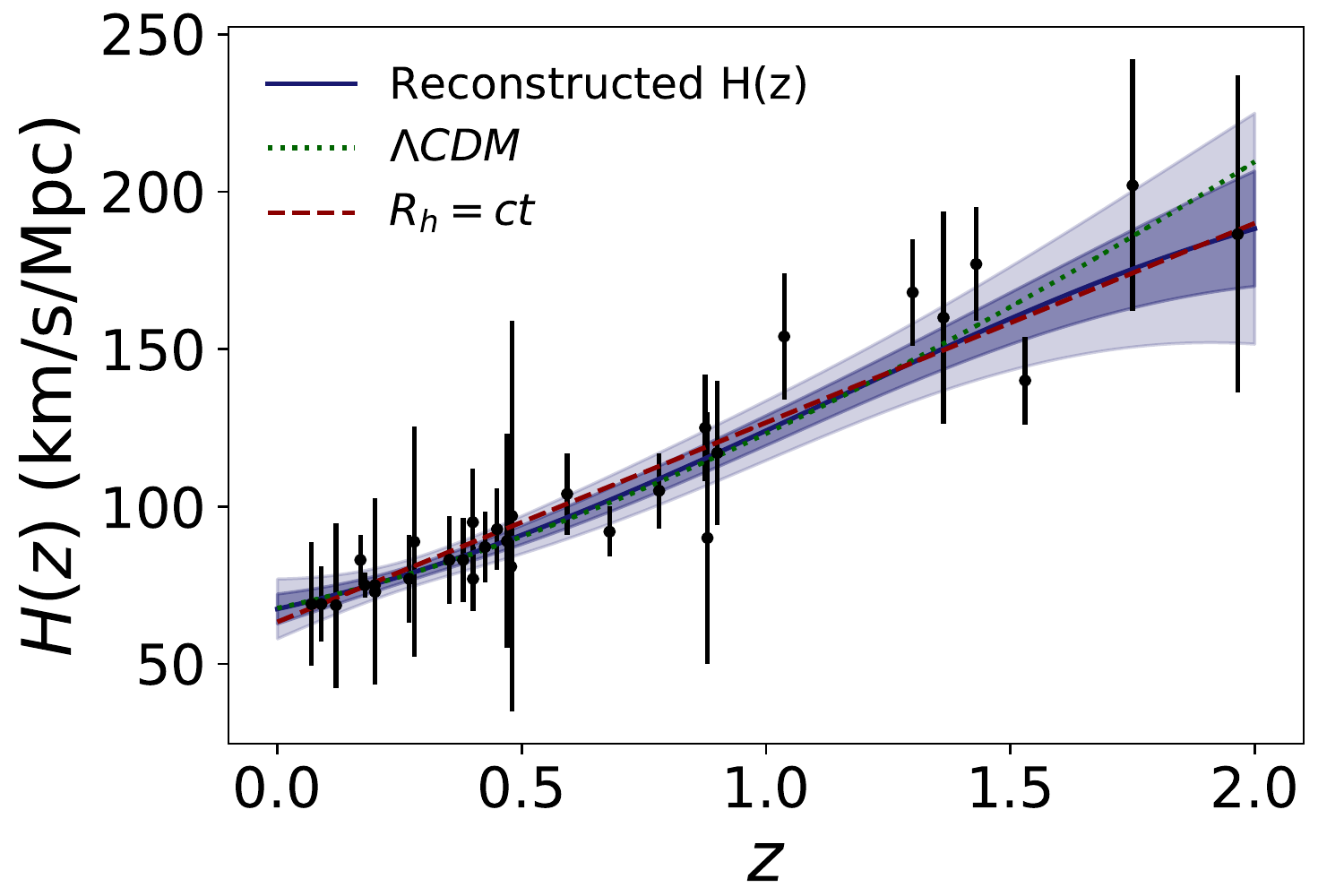}
\caption{Plot showing $H(z)$ chronometer data along with the best-fit  $\Lambda$CDM model and the $R_h = ct$ model, with best-fit parameters obtained using unbinned data. Also shown is   $H(z)$ reconstructed non-parametrically (along with $1\sigma$ and $2\sigma$ errors in reconstruction). The reconstruction was done with Gaussian Process Regression using  the {\tt GaPP} package.}
\label{fig:rh_c}
\end{figure}        

\section{Datasets and Analysis}
\label{sec:datasets}
 The $H(z)$ data from cosmic chronometers are obtained  by comparing relative ages of galaxies at different redshifts and is given by the following expression, assuming an FRW metric~\cite{Jimenez}: 
\begin{equation}
    H(z) = -\frac{1}{1+z}\frac{ dz}{d t}
    \label{eq:chrono}
\end{equation} 
Based on the measurements of the age difference, $\Delta t$, between two passively–evolving galaxies that are separated by a small redshift interval $\Delta z$, we can approximately calculate the value of $dz/dt$ from $\Delta z/ \Delta t$. This differential age method is much more reliable than a method based on an absolute age determination for galaxies, as absolute stellar ages are more vulnerable to systematic uncertainties than relative ages. 

Even though cosmic chronometers probe only the expansion history of the universe, they have been used for a variety of cosmological inferences, such as determination of $H_0$~\cite{ChenRatra,Amendola,Yang19,Haridasu2}, transition redshift from deceleration to acceleration~\cite{FarooqRatra,Jesus}, cosmic distance duality relation~\cite{Rana}, $\sigma_8$ estimation~\cite{Li}, dark energy equation of state~\cite{Ratra13,Moresco_2016}, etc.
The complete data set of 31 measurements of $H(z)$ at redshifts $0.07<z<1.965$ from cosmic chronometers is listed in Table \ref{tab:$H(z) data$}. This data set was obtained from the  compilation in Table III of Ref.~\cite{Li_2019}. A graphical summary of this unbinned data, along with the reconstructed $H(z)$ using GPR can be found in Fig.~\ref{fig:rh_c}. 

Although BS12 (and also Ref.~\cite{Lin}) has used $H(z)$ measurements from BAO to rule out $R_h=ct$ model, we have only used the Hubble parameter data  obtained from cosmic chronometers. This is due to   various  concerns regarding combining data from these two sources for parameter estimation within $\Lambda$CDM and for testing $R_h=ct$ universe~\cite{Zheng:2016jlq,Maier}.  One  problem  in using the BAO data for assessing the viability of an alternative to the $\Lambda$CDM model arises from the fact that measurement of the Hubble parameter from BAO requires the assumption of a particular cosmological model, unlike the model independent measurements of cosmic chronometers. All BAO measurements are scaled by the size of the sound horizon at the drag epoch, $r_s$. Computing the value of $r_s$ requires the assumption of a fiducial model. Most analyses which employ BAO measurements use the value of $r_s$ obtained using the $\Lambda$CDM model. This would induce a bias towards the $\Lambda$CDM model when comparing it with other models. Another concern is that one also needs to model the non-linear evolution of density and velocity fields, which are not model-independent~\cite{Maier,Meliaquasar}. Therefore, in MY18 and MM12, no BAO data was used, whereas both BAO and chronometer data was used in BS12. Accounting for all these problems we present our results for  model comparison  without the BAO data.

\begin{table}
    \centering
    \large
    \begin{tabular}{|c|c|c|c|}
        \hline
        $z$ & $H(z)$  & $\sigma$ & Ref.  \\
        &   (km/sec/Mpc) & (km/sec/Mpc) & \\ 
        \hline
        0.07 & 69 & 19.6 & \cite{Zhang_2014} \\
        0.09 & 69 & 12 & \cite{Simon_2005}\\
        0.12 & 68.6 & 26.2 & \cite{Zhang_2014}\\
        0.17 & 83 & 8 & \cite{Simon_2005}\\
        0.179 & 75 & 4 & \cite{Moresco_2011}\\
        0.199 & 75 & 5 & \cite{Moresco_2011}\\
        0.2 & 72.9 & 29.6 & \cite{Zhang_2014}\\
        0.27 & 77 & 14 & \cite{Simon_2005}\\
        0.28 & 88.8 & 36.6 & \cite{Zhang_2014}\\
        0.352 & 83 & 14 & \cite{Moresco_2011}\\
        0.3802 & 83 & 13.5 & \cite{Moresco_2016}\\
        0.4 & 95 & 17 & \cite{Simon_2005}\\
        0.4004 & 77 & 10.2 & \cite{Moresco_2016}\\
        0.4247 & 87.1 & 11.2 & \cite{Moresco_2016}\\
        0.4497 & 92.8 & 12.9 & \cite{Moresco_2016}\\
        0.47 & 89 & 34 & \cite{Ratsimbazafy_2017}\\
        0.4783 & 80.9 & 9 & \cite{Moresco_2016}\\
        0.48 & 97 & 62 & \cite{Stern_2010}\\
        0.593 & 104 & 13 & \cite{Moresco_2011}\\
        0.68 & 92 & 8 & \cite{Moresco_2011}\\
        0.781 & 105 & 12 & \cite{Moresco_2011}\\
        0.875 & 125 & 17 & \cite{Moresco_2011}\\
        0.88 & 90 & 40 & \cite{Stern_2010}\\
        0.9 & 117 & 23 & \cite{Simon_2005}\\
        1.037 & 154 & 20 & \cite{Moresco_2011}\\
        1.3 & 168 & 17 & \cite{Simon_2005}\\
        1.363 & 160 & 33.6 & \cite{Moresco_2015}\\
        1.43 & 177 & 18 & \cite{Simon_2005}\\
        1.53 & 140 & 14 & \cite{Simon_2005}\\
        1.75 & 202 & 40 & \cite{Simon_2005}\\
        1.965 & 186.5 & 50.4 & \cite{Moresco_2015}\\
        \hline
    \end{tabular}
    \caption{$H(z)$ data from cosmic chronometers  along with references to original sources.  This list was compiled from  Ref.~\cite{Li_2019}.}
    \label{tab:$H(z) data$}
\end{table} 

The first step in model comparison is to find the best-fit values of the free parameters in $\Lambda$CDM as well as the $R_h=ct$ universe model. This is obtained by minimizing the $\chi^2$ functional given by:
\begin{equation}
    \chi^2= \sum_{i=1}^N \left(\frac{H_i(z) -H^{\Lambda CDM/R_h=ct}(z,\theta)}{\sigma_i}\right)^2,
    \label{eq:chisq}
\end{equation}
where $H_i(z)$ indicate the various Hubble parameter measurements, $N$  is the total number of datapoints used, $H^{\Lambda CDM/R_h=ct}(z,\theta)$ encapsulates the relation for the Hubble parameter in $\Lambda$CDM and $R_h=ct$ cosmology; $\sigma_i$ denotes the error in $H(z)$; and $\theta$ denotes the parameter vector in the two models.


In the $R_h = ct$ model, $H(z)$ is given by: 
\begin{equation}
    H(z)=H_0(1+z),
\label{eq:rhct}    
\end{equation}
whereas for the the $\Lambda$CDM model, $H(z)$ is: 
\begin{equation}
    H(z)=H_0 \sqrt{\Omega_M(1+z)^3+(1-\Omega_M-\Omega_{\Lambda})(1+z)^2+\Omega_{\Lambda}}
    \label{eq:lcdm}
\end{equation}
where $\Omega_M$ and $\Omega_{\Lambda}$ are the density parameters of matter and the cosmological constant respectively. Note that for a flat $\Lambda$CDM model, $\Omega_{\Lambda}=1-\Omega_M$, which reduces the number of free parameters by one. For a flat $\Lambda$CDM model the equation would be - 
\begin{equation}
    H(z)=H_0 \sqrt{\Omega_M(1+z)^3+1-\Omega_M}
    \label{eq:flcdm}
\end{equation}

In both BS12 and MM13, a flat $\Lambda$CDM model was used for the model comparison. So in this work we will  stick to the flat case of the $\Lambda$CDM model ($\Omega_k = 0$) with $H(z)$ given from equation \ref{eq:flcdm}. \\

\rthis{Since Bayesian model comparison does not depend upon the best-fit values, we do not have to maximize any likelihood.
We only need to choose priors for the two models. For $\Lambda$CDM, we used two sets of  priors. The first set assumes a uniform distribution for $\Omega_M$ and $H_0$. For the second set of priors, we use the 2018 Planck cosmology determined best-fit parameters~\cite{Planck2018}, and choose Gaussian priors centered around these values. The  $R_h=ct$ universe
has only one free parameter, $H_0$ and we used the same (uniform) $H_0$ prior as in $\Lambda$CDM model. \footnote{\rthis{We do not use the Gaussian prior on the value of $H_0$ for $R_h=ct$ as the Planck 2018 results were obtained for the $\Lambda$CDM model, and there is no independent precise estimate of $H_0$ for $R_h=ct$ model.}}
A summary of all the priors used for model comparison for both the models can be found in Table~\ref{tab:priors}.} In this work, the Bayesian  evidence was computed using the  {\tt dynesty}~\cite{speagle_2019dynesty} package, which uses the nested sampling technique.


\section{Results}
\label{sec:results}
We now present our results   for model comparison using the chronometer dataset. We carried out two different analyses. The first analysis involves using the  unbinned data.
The second analysis involves reconstructing $H(z)$ using the non-parametric GPR method. \rthis{For each of these datasets, we used two different priors for $\Lambda$CDM, as outlined in the previous section}.
For this purpose, we  repeat the analysis done in BS12, wherein $H(z)$ is reconstructed at many values using GPR.
The GPR was done using the {\tt GaPP} software. 
This GPR reconstructed $H(z)$ for chronometers along with the original unbinned measurements  is shown in Fig.~\ref{fig:rh_c}, along with the best-fit  $\Lambda$CDM model and the $R_h = ct$ model.
For carrying out model comparison with GPR, we use 100 reconstructed measurements uniformly distributed between the lowest and highest available redshift.


\subsection{Model comparison using unbinned data}
Our model comparison results using unbinned analysis using both the prior choices are summarized in Table~\ref{tab:chrono}.
The summary of these results is  as follows. \rthis{When uniform priors for $\Lambda$CDM are chosen}, the Bayes factor (defined as ratio of Bayesian evidence for  $\Lambda$CDM model to $R_h=ct$) is close to one, and hence does not prefer any one model over the other.  \rthis{However, if we choose
Gaussian priors centered around Planck best-fit  values, then $\Lambda$CDM is very strongly favored over $R_h=ct$ using Jeffreys scale.} Therefore, we disagree with MM12 that $R_h=ct$ is favored, if you consider only chronometer data.

\subsection{Model Comparison using GPR data}
Our results for model comparison  using  data reconstructed with GPR  can be found in Tables~\ref{tab:chronowith_gp}.  The Bayes factor again marginally favors $\Lambda$CDM, \rthis{when uniform priors are used. When we use Planck based priors, then  $\Lambda$CDM is decisively favored over $R_h=ct$.}

Therefore, in summary we disagree with MS18 that $R_h=ct$ provides a better fit than the $\Lambda$CDM model, since no test provides a decisive evidence for either model and most tests strongly favor the $\Lambda$CDM model. At the same time we note that out $R_h=ct$ model cannot be currently ruled out using chronometers, \rthis{if we use uniform priors on $\Omega_M$ and $H_0$}.

\begin{table}[]
    \centering
    \begin{tabular}{|c|c|}
        \hline
         \multicolumn{2}{|c|}{$\Lambda$CDM - Uniform prior}  \\ \hline
         $\Omega_M$ & $\mathcal{U}(0,1)$  \\ \hline 
         $H_0$ & $\mathcal{U}(0,100)$ \\ \hline
         \hline
         \multicolumn{2}{|c|}{$\Lambda$CDM - Gaussian prior}  \\ \hline
         $\Omega_M$ & $\mathcal{N}(0.315,0.007)$  \\ \hline 
         $H_0$ & $\mathcal{N}(67.4,0.5)$ \\ \hline
         \hline
         \multicolumn{2}{|c|}{$R_h=ct$} \\ \hline
         $H_0$ & $\mathcal{U}(0,100)$ \\ \hline 
    \end{tabular}
    \caption{The priors used for the analysis. $\mathcal{U}(x,y)$ denotes a top-hat or a uniform prior between $x$ and $y$. $\mathcal{N}(x,y)$ denotes a Gaussian prior with a mean of $x$ and scale parameter of $y$. The Gaussian priors for $\Lambda$CDM are centered around the best fit values of the 2018 results of Planck collaboration \cite{Planck2018}, with the scale parameter equal to $1\sigma$ error of these results. The priors on $H_0$ are given in units of km/sec/Mpc}
    \label{tab:priors}
\end{table}

\begin{table*}
    \centering
    \begin{tabular}{|c|c|c|c|}
        \hline
        & $R_h=ct$  & $\Lambda$CDM  &  $\Lambda$CDM \\
          &  &(Uniform Prior) &  (Gaussian prior) 
          \\ \hline
         $\log Z$ & -128.0 & -129.3 & -123.9 \\ \hline 
         Bayes Factor & - & 0.3 & 60.0 \\ \hline 
         
         \hline
    \end{tabular}
    \caption{Comparison of Bayes factor for $R_h=ct$ and $\Lambda$CDM using unbinned measurements of  chronometer data listed in Table~\ref{tab:$H(z) data$}, using two different sets of priors in $\Lambda$CDM (cf. Table~\ref{tab:priors})
    $\log Z$ denotes the logarithm of the Bayesian evidence. The Bayes factor is defined as the ratio of the evidence of the $\Lambda$CDM model to the evidence for the $R_h=ct$ universe model. When uniform priors are used for $\Lambda$CDM,  the Bayesian evidence for the two models are almost identical, with no one model been preferred. When we used Gaussian priors centered on the  Planck best-fit values~\cite{Planck2018}, $\Lambda$CDM is very strongly preferred over  $R_h=ct$.}
    \label{tab:chrono}
\end{table*} 


    \begin{table*}
    \centering
    \begin{tabular}{|c|c|c|c|}
        \hline
        & $R_h=ct$  & $\Lambda$CDM  &  $\Lambda$CDM \\
          &  &(Uniform Prior) &  (Gaussian prior) 
          \\ \hline
         $\log Z$ & -277.7 & -277.3 & -270.8 \\ \hline 
         Bayes Factor & - & 1.6 & 992.3 \\ \hline

         \hline
    \end{tabular}
    \caption{Model Comparison tests using GPR measurements of  chronometer data listed in Table~\ref{tab:$H(z) data$}. The explanation of all the columns is the same as in Table~\ref{tab:chrono}.  When uniform priors are used, no one model is preferred, whereas $\Lambda$CDM is decisively favored if we use Gaussian  priors obtained from the 2018 Planck best-fit measurements~\cite{Planck2018}}
    \label{tab:chronowith_gp}
\end{table*}

\section{Diagnosis using $Om$ statistic}
\label{sec:om}
We now explore  if we can  distinguish between the two models using the two-point $Om(z_1,z_2)$ statistic between any two pairs of redshifts ($z_1$,$z_2$).
The $Om(z_1,z_2)$ statistic is defined as~\cite{Shafieloo}:
\begin{equation}
Om(z1,z2)= \frac{h^2(z_1)-h^2(z_2)}{(1+z_1)^3-(1+z_2)^3}
\label{eq:Omz}
\end{equation}
where $h(z)=H(z)/H_0$.
The $Om(z_1,z_2)$  statistic has been used to map out the expansion history of the universe and also as a null test of $\Lambda$CDM in a number of works~\cite{Shafieloo,Sahni,Qi,Cao,Fabris,Zheng:2016jlq}. For $\Lambda$CDM  model, $Om(z_1,z_2)$ has the remarkable property that it is independent of $z_1$ and $z_2$, and is equal to $\Omega_M$~\cite{Sahni}. Therefore, computing the $Om(z_1,z_2)$  using $H(z)$ measurements enables us to carry out a model independent test of $\Lambda$CDM and simultaneously obtain an estimate of $\Omega_M$. For $R_h=ct$ universe, $Om(z1,z2)$
 is given by
 \begin{equation}
 Om^{R_h=ct}= \frac{(1+z_1)^2-(1+z_2)^2}{(1+z_1)^3-(1+z_2)^3}
\end{equation}
Therefore for $R_h=ct$ model, $Om(z1,z2)$ is not a constant and is a function of $z_1$ and $z_2$.

\begin{figure}
        \includegraphics[width=0.5\textwidth]{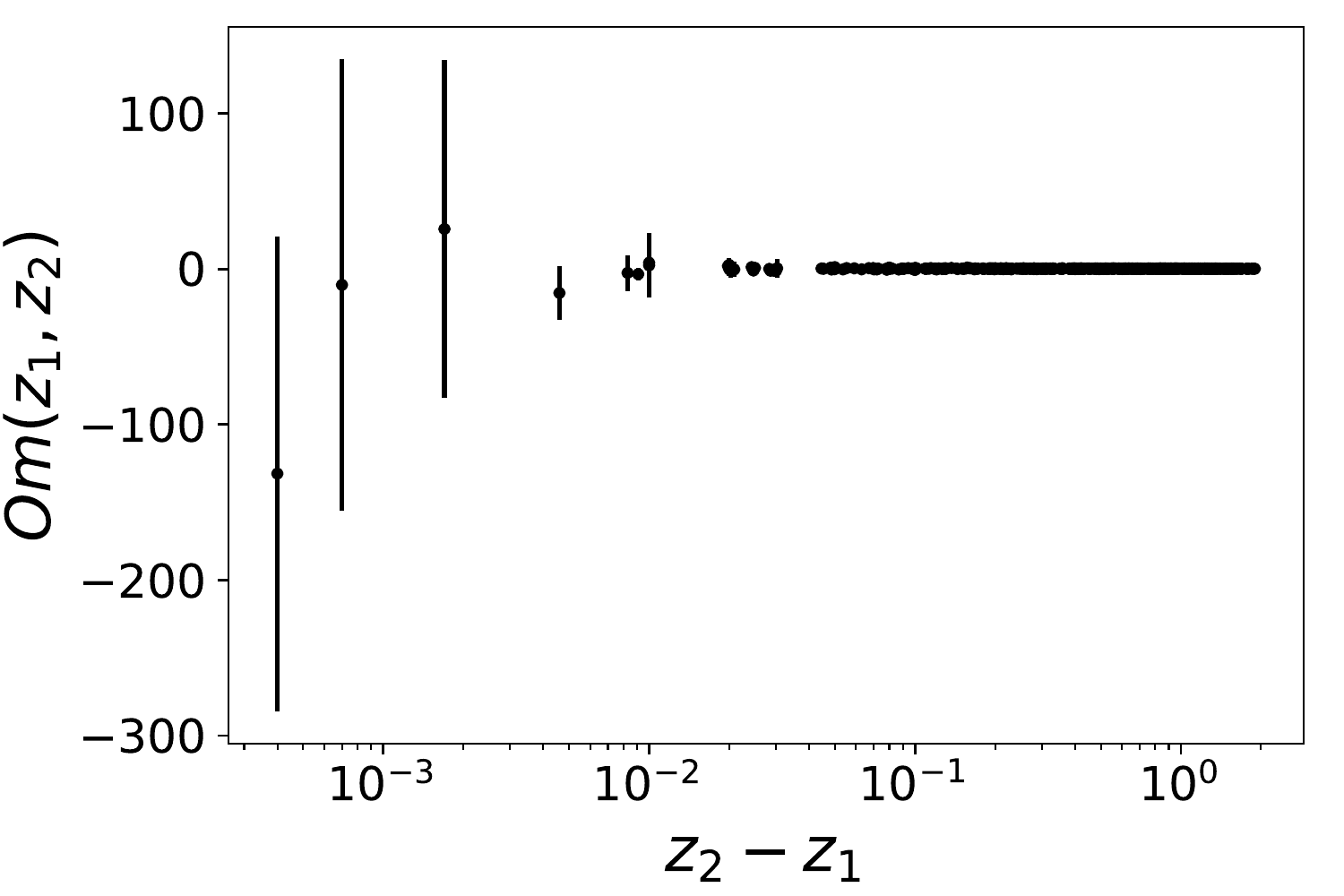}
        \caption{Plot of $Om(z_1,z_2)$ calculated from the chronometer data using equation \ref{eq:Omz}. These points were calculated from combinations of the 31 Hubble measurements in pairs, amounting to 465 points. The theoretical plots are not included because the theoretical curves cannot be distinguished at this scale.}
        \label{fig:omz1}
\end{figure}

From 31 $H(z)$ measurements, we obtain a total of $^{31}C_2$ or 465 $Om(z_1,z_2)$ data points. These data points can be found in Fig.~\ref{fig:omz1}.  The errors are obtained from Gaussian error propagation from the errors in $H(z_1)$ and $H(z_2)$.
As we can see, for low values of the redshift difference, the errors in $Om(z_1,z_2)$ are quite large,  and although they reduce with increasing $z_2-z_1$, they are usually of the same order as $Om(z_1, z_2)$.

For doing model comparison, we need to determine the total number of free parameters in  $\Lambda$CDM and $R_h=ct$. For $\Lambda$CDM, this  is equal to one, since $H_0$ is degenerate with $\Omega_m$, and choosing a different $H_0$ would lead to a different $\Omega_m$. However, irrespective of which value of $H_0$ is used, $Om(z1,z2)$ would  be a constant, independent of the redshift difference. Since $Om(z1,z2)$  is constant for $\Lambda$CDM model, the best-fit maximum likelihood estimate would just be the weighted mean of all the $Om(z1,z2)$ measurements.
For $R_h=ct$, the only free parameter  would be $H_0$, since varying $H_0$ would  vertically re-scale the whole plot by a constant offset.

\begin{figure}
        \includegraphics[width=0.5\textwidth]{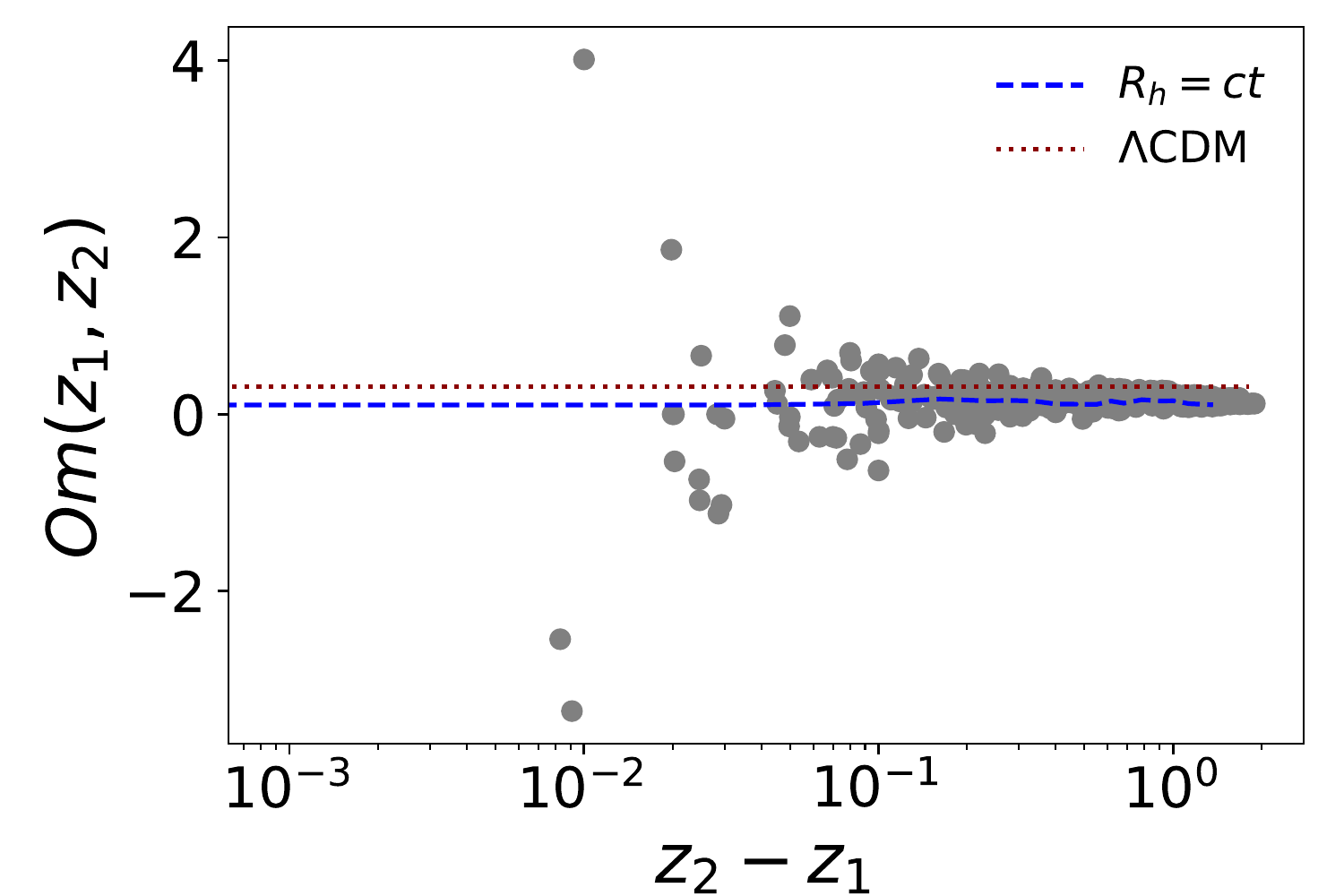}
        \caption{This plot shows the theoretical curves for  the best-fit $\Lambda$CDM and $R_h=ct$ models , along with the data for  $Om(z_1,z_2)$ (grey points). For this plot, four  $H(z)$ points with the largest error bars  have been removed for which the value of $Om(z_1,z_2)$ was large. Since this plot is for illustrative purposes, we have also removed the error bars in the $Om(z_1,z_2)$ values for brevity. The actual error bars are much larger than the differences between the two models. Therefore, it is not possible to distinguish between the two models using $Om(z_1,z_2)$ measurements.}
        \label{fig:omz2}
\end{figure}

For $\Lambda$CDM, we get  $$ \chi^2/dof = 185.4/350$$ and for $R_h=ct$ we get $$\chi^2/dof = 185.2/350$$
For doing this fit, we removed  four $H(z)$ points with the largest error bars. So the total number of $Om(z_1,z_2)$ data points used for doing the fits is equal to 351. As we see, both the $\chi^2/dof$ values are smaller than one and are very close to each other making the $Om(z_1,z_2)$ ineffective for this model comparison. For illustrative purposes, we show this best-fit along with some of the $Om(z_1,z_2)$ (after removing the error bars) in Fig.~\ref{fig:omz2}. Therefore, it is not possible to distinguish between the two models using current $H(z)$ chronometer data.

\section{Conclusions}
\label{sec:conclusions}
In this work we try to independently assess the viability of $\Lambda$CDM vs $R_h=ct$ universe using only $H(z)$ measurements from cosmic chronometers to resolve conflicting claims between two groups of authors. In 2012, Bilicki and Seikel~\cite{Seikel_2012_rhct} claimed using $H(z)$ measurements from chronometers and BAO, that $R_h=ct$ model is conclusively ruled out. This was contested by Melia and collaborators ~\cite{Maier,Meliachrono}, who showed using $H(z)$ measurements from chronometers that $R_h=ct$ universe  is favored over $\Lambda$CDM.
They also pointed out BAO measurements cannot be used to test $R_h=ct$ models, since the BAO $H(z)$ measurements implicitly assume $\Lambda$CDM. A few  other works~\cite{Shafer,Lin,Haridasu1,Tu2019} also found that type 1a SN, $H(z)$ measurements  from chronometers, and BAO rule out $R_h=ct$ model.

In order to settle the conflicting results between the above two groups of authors, we considered  measurements from  only chronometers (to emulate the analysis in Ref.~\cite{Maier, Meliachrono}).) We did not consider the BAO measurements, given the circularity involved in using them for testing non-$\Lambda$CDM universes~\cite{Maier,Meliachrono}.
We carried out model comparison using both the unbinned  data, and also by doing a non-parametric reconstruction using GPR. \rthis{To carry out model comparison, we used a Bayesian model comparison technique by computing the Bayes factor between the two models. We used two different priors for the $\Lambda$CDM: a uniform prior over a wide parameter range, and also Gaussian priors centered around the 2018 Planck best-fit $\Lambda$CDM cosmology. A summary of these priors used can be found in Table~\ref{tab:priors}}.

Our results for both these priors and datasets can be found in Tables~\ref{tab:chrono} and ~\ref{tab:chronowith_gp}.  
When we use a uniform prior, the difference in significance between the two models is negligible, using both the datasets.  \rthis{However, for the priors  centered around the Planck 2018 best-fit $\Lambda$CDM values, we find that $\Lambda$CDM is very strongly/decisively favored over $R_h=ct$ for the unbinned/GPR reconstructed datasets.}
Therefore, we conclude that using  the chronometer $H(z)$ data, $R_h=ct$ model is not preferred over $\Lambda$CDM.

We also investigated if the $Om$ statistic, calculated using redshift pairs, which has been used in previous literature for testing $\Lambda$CDM model~\cite{Shafieloo,Sahni}, can be used to discriminate between the two models.
Unfortunately, the current error bars in $Om(z_1,z_2)$ estimated using chronometer $H(z)$
data are too large to enable a robust model comparison.

Therefore, in summary, we disagree with the claims in both  Ref.~\cite{Seikel_2012_rhct} and Ref.~\cite{Maier, Meliachrono}, and conclude that neither model is ruled out or decisively favored using only $H(z)$ measurements with chronometers, \rthis{if we use uniform priors on parameters of both models.}
A more acid test would be using CMB and other large scale structure based tests in a theory-independent fashion.

\section*{Acknowledgements}
We are grateful to Fulvio Melia and Varun Sahni   for useful correspondence \rthis{and the anonymous referee for constructive feedback on the manuscript.}
\bibliography{ref}

\end{document}